\documentclass[twocolumn,preprintnumbers,amsmath,amssymb,superscriptaddress]{revtex4-2}

\usepackage{mathrsfs}
\usepackage{amssymb}
\usepackage{graphicx}
\usepackage{dcolumn}
\usepackage{bm}
\usepackage{textcomp}
\usepackage{amsmath}
\usepackage[titletoc]{appendix}
\usepackage{accents}
\usepackage{ntheorem}
\usepackage{color}

\newtheorem{lemma}{Lemma}

\newtheorem{theorem}{Theorem}

\newcommand\ket[1]{\ensuremath{|#1\rangle}}

\newcommand\oprod[2]{\ensuremath{|#1\rangle\langle#2|}}
\newcommand\mean[1]{\ensuremath{\langle #1\rangle}}

\newcounter{RomanNumber}

\makeatletter
\def\widebar{\accentset{{\cc@style\underline{\mskip10mu}}}}
\def\Widebar{\accentset{{\cc@style\underline{\mskip8mu}}}}
\makeatother

\begin{document}

\title{Composable security for practical quantum key distribution with two way classical communication}

\author{Cong Jiang}
\affiliation{State Key Laboratory of Low Dimensional Quantum Physics, Department of Physics, Tsinghua University, Beijing 100084, P.~R.~China}
\affiliation{Jinan Institute of Quantum Technology, Jinan, Shandong 250101, P.~R.~China}

\author{Xiao-Long Hu}
\affiliation{School of Physics, State Key Laboratory of Optoelectronic Materials and Technologies, Sun Yat-sen University, Guangzhou 510275, China}

\author{Zong-Wen Yu}
\affiliation{State Key Laboratory of Low Dimensional Quantum Physics, Department of Physics, Tsinghua University, Beijing 100084, P.~R.~China}
\affiliation{Data Communication Science and Technology Research Institute, Beijing 100191, P.~R.~China}

\author{Xiang-Bin Wang}\email{Corresponding author: xbwang@mail.tsinghua.edu.cn}
\affiliation{State Key Laboratory of Low Dimensional Quantum Physics, Department of Physics, Tsinghua University, Beijing 100084, P.~R.~China}
\affiliation{Jinan Institute of Quantum Technology, Jinan, Shandong 250101, P.~R.~China}
\affiliation{Shanghai Branch, CAS Center for Excellence and Synergetic Innovation Center in Quantum Information and Quantum Physics, University of Science and Technology of China, Shanghai 201315, P.~R.~China}
\affiliation{ Shenzhen Institute for Quantum Science and Engineering, and Physics Department, Southern University of Science and Technology, Shenzhen 518055, China}
\affiliation{Frontier Science Center for Quantum Information,
Beijing, China.}
\begin{abstract}
	We present  methods to strictly calculate the finite-key effects in quantum key distribution (QKD) with error rejection through two-way classical communication (TWCC) for the sending-or-not-sending twin-field protocol. Unlike the normal QKD without TWCC, here the  probability of tagging or untagging for each two-bit random group is not independent. We rigorously solve this problem by imagining a virtual set of bits where every bit is independent and identical. We show the relationship between the outcome starting from this imagined set containing independent and identical bits and the outcome starting with the real set of non-independent bits. With explicit formulas, we show that simply applying Chernoff bound in the calculation gives correct key rate, but the failure probability changes a little bit.
\end{abstract}


\maketitle
\section{Introduction}
As a crucially important issue of practical quantum key distribution (QKD)~\cite{bennett1984quantum,pirandola2020advances,xu2020secure,gisin2002quantum,
scarani2009security,hwang2003quantum,wang2005beating,lo2005decoy,
lo2012measurement,braunstein2012side,
yin2016measurement,liao2017satellite,boaron2018secure,lu2018overcoming}, 
the finite-key effect has been extensively studied in the past~\cite{muller2009composability,renner2005security,konig2007small,tomamichel2012tight,curty2014finite,lim2014concise,jiang2019unconditional,maeda2019repeaterless,
jiang2020zigzag}. These studies show that secure QKD in practice is possible. However, the finite-key study on QKD with two-way classical communication (TWCC)~\cite{chau2002practical,gottesman2003proof,wang2004quantum,kraus2007security} is rare. Through TWCC, one can take error rejection by parity check on those randomly grouped two-bit pairs and reduce the bit-flip errors effectively. Importantly, this raises the fault-tolerance performance of QKD~\cite{gottesman2003proof,chau2002practical,wang2004quantum,kraus2007security}.  Given the potential importance of TWCC for QKD, a robust theory for finite-key effects of QKD with TWCC shall be especially useful. Here we present such a theory. 

{\em Short review and the major  problem on finite-key effects with TWCC.}
The central idea in TWCC is to take {\em bit-flip error rejection} through random grouping and parity check which requires two-way classical communication. In the standard TWCC, Alice and Bob shall randomly group their bits of $Z$ basis (coding basis) two by two and obtain many pairs of bits (two-bit groups). They perform parity check on each pair and discard those pairs with different parity values while keeping one bit of any pair with the same parity values of two sides. In this way, the bit-flip error rate will be effectively reduced.

However, if we apply TWCC to the decoy-state method, we need to verify the number of untagged pairs containing two untagged bits. Each bits are not independent on tagging or untagging. A strict treatment of this is needed and here we present such a strict treatment. Our study show that if one simply applies Chernoff bound for this step, the key rate from the calculation is still correct, though the failure probability changes a little bit.
Surely, the study for finite key effects of TWCC with strict bounds is crucially important for security of fault tolerant QKD. However, so far study towards this end is rare. Ref.~\cite{jiang2020zigzag} has studied finite key size~\cite{muller2009composability,renner2005security,konig2007small,tomamichel2012tight} to bound the phase-flip errors in sending-or-not-sending (SNS)~\cite{wang2018twin} protocol of twin-field (TF) QKD~\cite{lu2018overcoming} with TWCC~\cite{xu2020sending}. Here,we present a simple and rigorous study for both the upper bound of phase error rate and lower bound for the number of untagged bits after error rejection. With these, we calculate key rate strictly with the composable security.    

We associate sifted bits in the real protocol mathematically with a virtual set of independent identical bits. We seek conditions with high probability when the outcome of the virtual set is worse than that of the real bits after error rejection. Based on this idea, we present mathematical formulas pointing directly to the lower  bound of the number of untagged bits after error rejection in the real protocol, with an explicitly known small failure probability. Using our method, calculating the number of untagged bits after error rejection in the real bits is transformed to calculating the value with a virtual set containing independent and identical bits, and hence the strict bound values are easily obtained with existing methods, such as the Chernoff bound~\cite{chernoff1952measure}.

This paper is arranged as follows: In Sec.~\ref{mainr}, we show the theorems on how to make a strict and efficient method to estimate parameter values after error rejection. In Sec.~\ref{twccsns} and Sec.~\ref{opersns}, we show how to apply our results to the SNS protocol with the standard TWCC, OPER and AOPP methods. We then take numerical calculations on the SNS protocol with the standard TWCC method and its variants.

\section{Mathematical model with white balls and black balls}\label{mainr}
We extract the question into the following mathematical model: {Set {$\mathcal U$} contains $N$ balls, some of them are white and some of them are black. There are two kinds of sets. In one type of set, $k$ of them are white and $N-k$ of them are black. In another type of set, every ball has independent and identical probability to be white.} After random grouping, at least how many pairs containing two white balls are created with a certain failure probability ?  (We assume $N$ to be an even number throughout this paper, and we shall also assume the number of elements of any subset of {$\mathcal U$} to be an even number if we take random grouping to the elements in the subset.)

For clarity, we list our important notations first.

{\bf Notation 2.1} $[k,N]$: a set  containing $N$ balls, among which there are $k$ white balls and $N-k$ black balls.
 
{\bf Notation 2.2} $[p_u,N]_{iid}$: a set  containing $N$ balls, where every ball has an independent and identical probability $p_u$ to be white, and probability $1-p_u$ to be black. 

{\bf Notation 2.3}
$n_{\alpha\beta}$: the observed number of $\alpha\beta$-pairs after random grouping to the corresponding set of balls. The $\alpha\beta$-pair can be a $ww$-pair that contains two white balls or a $wb-$pair that contains a white ball and a black ball. We shall study the failure probability of {$n_{\alpha\beta}\ge \underline n_{\alpha\beta}$, where $\underline n_{\alpha\beta}$} is a specific bound.

{\bf Notation 2.4} $\varepsilon(\underline n_{\alpha\beta} | {\mathcal U})$ : probability that the number of $\alpha\beta$-pairs is less than $\underline n_{\alpha\beta}$ after random grouping to balls initially in set $\mathcal{U}$. The set can be $[k,N]$ or $[p_u,N]_{iid}$.

According to {\bf Notation 2.2}, set $[p_u,N]_{iid}$ can be regarded as the probability distribution over set {$[m,N]$}. Explicitly, the probability on {$[m,N]$ is:
\begin{equation}\label{pm1}
\tilde p(m) =C_N^m p_u^m (1-p_u)^{N-m}.
\end{equation}}

With these notations, we now present our major mathematical conclusions below. We shall show the proofs in Appendix~\ref{lemma11}.

\begin{lemma}\label{ll1}
	For sets  $W_1=[k_1,N]$, $W_2=[k_2, N]$,
	the inequality
	\begin{equation}\label{lemin1}
	\varepsilon(\underline{n}_{ww}|W_1)\le \varepsilon(\underline{n}_{ww}|W_2)
	\end{equation}
	always holds for whatever non-negative integer $\underline{n}_{ww}$
	provided that $k_1 \ge k_2\ge 0$.
\end{lemma}

{\bf{Result 2.1}}: If we take random pairing to  set $[k,N]$, we can calculate the lower bound of the number of $ww$-pairs by the Chernoff bound or any other tail bounds assuming the independent probability of $k/N$ for every ball, with the failure probability multiplied by $2$. Mathematically:
\begin{equation}
\varepsilon(\underline n_{ww}|[k,N])\le \epsilon=2\varepsilon( \underline n_{ww}|[p_u=k/N,N]_{iid}).
\end{equation}
	
{This result shows that we can simply regard the set $[k,N]$ as the set $[p_u=k/N,N]_{iid}$ in calculating $\underline n_{ww}$ for the input set $[k,N]$, we only need multiply the failure probability by a factor 2. For set $[p_u,N]_{iid}$, we can use the existing ways such as the Chernoff bound or the numerical bound calculating the lower bound here because every ball in set $[p_u,N]_{iid}$ is independent and identical. In particular
	\begin{equation}\label{fapro}
	\epsilon = 2 \sum_{l=0}^{\underline n_{ww}-1} p_u^{2l}(1-p_u^2)^{N/2-l}C_{N/2}^{l}.
	\end{equation} 
	
More conveniently, we can relate the failure probability with the standard value from a Binomial distribution $B(M,p)$ where there are $M$ elements and each element has an independent and identical probability $p$ to be $"1"$. If we denote $\xi_L(\underline x; p,M)$ as the probability of obtaining less than $\underline x$ "1" from a binomial distribution set $B(M,p)$, we can formulate
\begin{equation}\label{bionom}
\xi_L(\underline x; p,M)=\sum_{l=0}^{\underline x-1}p^{l}(1-p)^{M-l}C_{M}^{l}.
\end{equation}
Eq.(\ref{fapro}) can be written in
	\begin{equation}
	\epsilon = 2 \xi_L(\underline n_{ww};p_u^2,N/2).
	\end{equation}
	
A detailed  proof for the more general form of {\bf Result 2.1}, Theorem 1, is given in Appendix~\ref{lemma11}. But here we can show it in a simple way:}

{Proof of Result 2.1: Given the input set $[p_u=k/N, N]_{iid}$, we denote $n_w$ to be the observed number of white balls in set $[k/N,N]_{iid}$. We define 
	\begin{equation}\label{probabA}
	P_A =\sum_{n_w\le k}\tilde p(n_w)
	\end{equation}
as the  probability for $n_w\le k$ and $P_B= \sum_{n_w>k}\tilde p(n_w)$ as the probability for $n_w>k$. Here $\tilde p(n_w)$, as defined in Eq.(\ref{pm1}) is {the probability of observing $n_w$} white balls in set $[k/N,N]_{iid}$. Strictly~\cite{lord2010binomial}, $P_A=1/2+\delta_0$ is a bit larger than 1/2 and $P_B = 1/2-\delta_0$ is a bit smaller than 1/2. Define
\begin{equation}\label{epsa}
\kappa_A = \sum_{n_w\le k}\tilde p(n_w)\varepsilon(\underline{n}_{ww}|[n_w,N])
\end{equation} 
and $\kappa_B = \sum_{n_w> k}\tilde p(n_w)\varepsilon(\underline n_{ww}|[n_w,N])$. 

Obviously, \begin{equation}\label{aka3}
\varepsilon( \underline n_{ww}|[k/N,N]_{iid})= \kappa_A + \kappa_B\ge \kappa_A, 
\end{equation}
Applying Lemma 1, we have 
\begin{equation}\label{epsa1}
\kappa_A \ge \left(\sum_{n_w\le k}\tilde p(n_w) \right) \varepsilon(\underline n_{ww}|[k,N])= P_A \varepsilon(\underline n_{ww}|[k,N])
\end{equation} because no $n_w$ {inside the summation 
of Eq.(\ref{epsa}) can be larger than $k$. By Lemma 1, every term $\varepsilon(n_{ww}|[n_w,N])$ inside the summation in the right hand side of Eq.(\ref{epsa1}) has to respect $\varepsilon(n_{ww}|[n_w,N])\ge \varepsilon(n_{ww}|[k,N])$.} We have used the definition of $P_A$ in Eq.(\ref{probabA}) in the second equality above.

 {Combining Eq.(\ref{aka3}) and Eq.(\ref{epsa1}) we obtain
 \begin{equation}\label{aka5}
\varepsilon( \underline n_{ww}|[k/N,N]_{iid}) 
\ge P_A \varepsilon(\underline n_{ww}|[k,N]), 
\end{equation} }
which concludes
 \begin{equation}\label{aka6}
 \begin{split}
\varepsilon(\underline n_{ww}|[k,N])&\le \frac{\varepsilon( \underline n_{ww}|[{k}/{N},N]_{iid})}{P_A} \\
&\le  2\varepsilon(\underline n_{ww}|[{k}/{N},N]_{iid})\\
&= \epsilon.
\end{split}
\end{equation} 
We have used the fact that $P_A$ is a bit larger than $1/2$ above.
This completes our proof. Surely, {if $k$ is the lower bound value rather than the exact value for the number of white balls in set $[k,N]$,} {\bf Result 2.1} still holds because of Lemma 1. 

Actually, we are not limited to use the specific setting of $k/N$. We have more general result presented as }
\begin{theorem}\label{th11}
The inequality	
	\begin{equation}\label{theo1u}
\varepsilon(\underline n_{ww}|[k_u,N])\le \frac{\varepsilon(\underline{n}_{ww}|[p_u,N]_{iid})}{\gamma_{iid}},
	\end{equation}
always holds with whatever non-negative integer $k_{u}$, $\underline {n}_{ww}$, and whatever probability value $p_u$. Here 
\begin{equation}
\gamma_{iid}=\sum_{k=0}^{k_{u}} \tilde p(k),
\end{equation}
where $\tilde p(k)$ is defined in Eq.~\eqref{pm1}
\end{theorem}

\section{TWCC-SNS}\label{twccsns}
As an important variant of twin field QKD~\cite{lu2018overcoming}, the SNS protocol~\cite{wang2018twin} together with its modified protocols~\cite{yu2019sending,hu2019general,xu2020sending,jiang2020zigzag} have attracted many attentions due to its large noise tolerance and high key rate.  Moreover, the SNS protocol has a unique advantage that the traditional decoy-state method directly applies, which makes the finite-key analysis very efficient. The SNS protocol has been experimentally demonstrated in proof-of-principle in Ref.~\cite{minder2019experimental}, and realized in real optical fiber with the finite-key effects taken into consideration~\cite{chen2020sending,liu2019experimental}. Notably, the SNS protocol has been experimentally demonstrated over 509 km optical fiber~\cite{chen2020sending} which is the longest secure distance of QKD in optical fiber.

Here, applying our mathematical results above, we shall take the strict bound calculation for the finite key effects on the SNS protocol with standard TWCC method (TWCC-SNS)~\cite{xu2020sending} in the post data processing, which can also be applied to other protocols of decoy-state method~\cite{hwang2003quantum,wang2005beating,lo2005decoy,tamaki2012phase2,wang2013three,
xu2014protocol,yu2015statistical,zhou2016making,jiang2021higher}, obviously. In the standard TWCC-SNS \cite{xu2020sending}, for any pair, if both sides observed the same parity value no matter it is odd or even, we shall take one bit from this pair for final key distillation. We shall directly apply the method above for both the number of un-tagged bits and  phase-flip errors after error rejection, as requested for final-key calculation.

After light-pulse transmission, post selection and error test in the protocol, there are $n_t$ remaining bits for $Z$-basis which will be used for the final key distillation. We denote these $n_t$ bits by set $W$. Suppose there are $n_1$ untagged bits in set $W$. We denote $W_u$ for the set of these $n_1$ untagged bits. By the decoy-state analysis we can verify the lower of $n_1$, say, $\underline{n_1}$. We define the untagged pair as a pair that contains two bits from set $W_u$ after random grouping. We also name an untagged pair as a $uu$ pair. 

Suppose there are $n_{uu}$ untagged pairs after random grouping to all bits in set $W$. Regarding bits in set $W_u$ as the white balls in our Result 2.1, we can immediately lower bound the number of $uu$ pair $n_{uu}$ by
\begin{equation}
n_{uu} \ge \underline n_{uu},
\end{equation}
except for a probability $\epsilon_{twcc}$,
\begin{equation}
\epsilon_{twcc}=2\xi_L(\underline n_{uu};\frac{\underline{n_1}^2}{n_t^2},\frac{n_t}{2})
\end{equation}
{where $\xi_L$ is defined in Eq.(\ref{bionom})}

We use notation $U$ for the set of these $n_{uu}$ pairs, and notation $V$ for those $2n_{uu}$ untagged bits which form these $n_{uu}$ untagged pairs in set $U$. Suppose there are $m_{v_e}$ phase errors in set $V$. As shown below in Remark 3.1, set $V$ is a random subset of set $W_u$. Therefore the upper bound value $m_{v_e}$ can be faithfully and efficiently estimated by decoy-state analysis.

Since there is no bit-flip error for untagged bits in SNS protocol, so all those $n_{uu}$ untagged pairs will pass the parity check for sure and they will contribute $n_{uu}$ bits after discarding one bit in each pair. Our task now is to faithfully upper bound the phase-flip error rate of $n_{uu}$ survived untagged bits after error rejection.

There are two kinds of untagged pairs: a phase-error pair that contains one phase error only and a perfect pair that either contains $0$ phase error or $2$ phase errors.  As shown in the prior art works~\cite{gottesman2003proof,chau2002practical}, a phase error pair will produce a bit with one phase-error for sure after error rejection and the perfect pairs will not produce any phase error after error rejection step. 

Given the number of phase errors $m_{v_e}$ in set $V$, we have the following equation for the number of phase-error pairs
\begin{equation}\label{mie}
n_{Ie}=m_{v_e} - 2 n_{ee}
\end{equation}
where $n_{Ie}$ is the number of phase-error pairs and $n_{ee}$ is the number of pairs containing two errors. The formula above is based on the simple fact: those $n_{ee}$ pairs containing two phase errors have consumed $2n_{ee}$ phase errors in set $V$, each of the remaining $m_{v_e}$ phase errors will only be paired with a perfect bit.

Immediately we have
\begin{equation}\label{strict3}
n_{Ie}\le m_{v_e}.
\end{equation}
If we use this strict bound, we don't have to consider any statistical fluctuation at this step, because this is the worst-case result already, whose failure probability is $0$. Though we have no way to know the exact value of $m_{v_e}$, we can upper bound it by $m_{v_e} \le \bar m_{v_e}$ by decoy state analysis, given that set $V$ is a random subset of set $W_u$ as shown below.

{Remark 3.1:  We consider the following game: Clare initially keeps those bits in set $W_u$ and Bob keeps all the other bits.  Clare randomly permutes all those bits privately and after that he places each bit inside a sealed envelope and passes all envelopes to Bob.  Bob is not allowed to use any bit information inside the envelope. Under such a condition, in whatever way Bob may take( including the way that he uses additional bits kept by himself), he has no way to produce a subset of $W_u$ that is not a random subset of $W_u$. Therefore, subset $V$ above can only be a random subset of $W_u$. The mathematical conclusion does not depend on who takes random permutation or whether to place the bits inside envelopes. It simply means that, if initially all bits in set $W_u$ are randomly permuted, no mater who creates the randomness, any subset of bits must be a random subset of $W_u$ provided that bits in the subset are chosen in a way independent of bit value or phase error information of any bits.}

Numerical simulation shows that satisfactory key rate can be obtained by the simple and strict bound of Eq.(\ref{strict3}).

Consequently, we can calculate the final key rate (per sent pulse) of SNS protocol\cite{wang2018twin} with standard TWCC by formula
\begin{equation}\label{r2}
\begin{split}
R=&\frac{1}{N_{tol}}\{{\underline n}_{uu}[1-H({\bar e}_{Ap})]-f[n_{t1}H(E_1)+n_{t2}H(E_2)\\
&+n_{t3}H(E_3)]-\log_2 \frac{2}{\varepsilon_{sec}}-2\log_2\frac{1}{\sqrt{2}\varepsilon_{PA}\hat{\varepsilon}}\}
\end{split}
\end{equation}
where $N_{tol}$ is the total number of pulse pairs sent by Alice and Bob, and $n_{t1},n_{t2},n_{t3}$ are number of survived bits from different kinds of pairs, ${\bar e}_{Ap}=\bar m_{v_e}/\underline n_{uu}$ is the upper bound of phase-flip error rate for those survived untagged bits after error rejection. They are distinguished by a pair containing one bit value 1 and one bit value 0, a pair containing two bit value 0, and a pair containing two bit value 1. $E_1,E_2,E_3$ are bit-flip errors of each pair. Tailing term of
$-\log_2 \frac{2}{\varepsilon_{sec}}-2\log_2\frac{1}{\sqrt{2}\varepsilon_{PA}\hat{\varepsilon}}$ is the additional cost for security with finite size as shown in Ref.~\cite{curty2014finite,jiang2019unconditional}.

With the key rate formula~\eqref{r2}, the protocol is $\varepsilon_{tol}=\varepsilon_{cor}+\varepsilon_{sec}$-secure~\cite{tomamichel2012tight,curty2014finite} where 
\begin{equation}
\varepsilon_{sec}=2\hat{\varepsilon}+\varepsilon_{PA}+4\sqrt{\varepsilon_{e}+\varepsilon_{n_1}},
\end{equation} 
$\varepsilon_{e}$ is the failure probability of the estimation of phase-flip error rate, $\varepsilon_{n_1}$ is the failure probability of the estimation of the number of untagged bits in Eq.~\eqref{r2}, $\varepsilon_{cor}$ is the failure probability of error correction, $\varepsilon_{PA}$ is the failure probability of privacy amplification, and $\hat{\varepsilon}$ is the coefficient while using the chain rules of max- and min- entropy~\cite{jiang2019unconditional}.
\section{Odd-parity error rejection and actively odd-parity pairing}\label{opersns}
By the similar idea, we can strictly take the effects of finite data size to calculate the key rate of SNS protocol by other TWCC method such as the odd-parity error rejection (OPER) and the actively odd-parity pairing (AOPP)~\cite{xu2020sending,jiang2020zigzag} which can further improve the key rate significantly. 
\subsection{More mathematical results with white balls and black balls}
We need some additional mathematical results. The details of the proof are shown in Appendix~\ref{lemma11}.

{\bf Result 4.1:} After random pairing to balls in set $[k,N]$ that contains $N$ balls, the following inequality always holds
\begin{equation}
\varepsilon(\underline n_{wb}|[k,N]) \le 2\xi_L(\underline n_{wb}+1; 2p_u(1-p_u), N/2)
\end{equation}
provided that $k\le N/2$ and $p_u=k/N$. {Here functional $\xi_L$ is defined in Eq.(\ref{bionom}).}

\subsection{OPER-SNS}
We name the SNS protocol~\cite{wang2018twin} with OPER \cite{xu2020sending,jiang2020zigzag} in the post data processing as OPER-SNS. Again, we need bound values of two quantities after error rejection: the number of survived untagged bits from odd parity pairs and their phase error rate.

We can upper bound the number of phase errors in those survived untagged bits after OPER, and denote the upper bound number as $\bar{M}_{oper}$. This can be done by the zigzag approach~\cite{jiang2020zigzag} basing on the quantum de Finetti theorem~\cite{renner2005security,renner2007symmetry}. For completeness, we also present the details of Zigzag approach in Appendix~\ref{oper} of this work, in a readable way.

The number of survived untagged bits after OPER is just the number of odd parity untagged pairs $n_{oper}$. Its lower bound $\underline n_{oper}$ can be calculated by applying {\bf Result 4.1}, with the input of values $n_t,\underline{n_{01}}, \underline{n_{10}}$ before TWCC. Here $n_t$ is the total number of effective bits in
$Z$ basis, $\underline{ n_{01}}, \underline {n_{10}}$ are lower bounds number of untagged bits with bit value $0$ and $1$ respectively.

Since in the zigzag approach, we only need the survived $\underline{n}_{uu}$ untagged pairs after random grouping the $n_t$ bit that containing $\underline{n_1}$ untagged bits. We only need to study those $\underline{n}_{uu}$ untagged pairs in the calculation of $\underline n_{oper}$, too. Consider set $Y$ containing {those} $2\underline{n}_{uu}$ bits that formed those $\underline{n}_{uu}$ untagged pairs, {among which there are at least} $\underline{n_{01}^\prime}$ untagged bits $0$ and $\underline{n_{10}^\prime}$ untagged bits $1$. Let $n_{min}=\min(\underline{n_{01}^\prime},\underline{n_{10}^\prime})$.
{Relating $n_{min}$ to the number of white balls in {\bf Result 4.1},} we have $n_{oper}\ge\underline{n}_{oper}$ with a failure probability $\epsilon_{oper}$
and   
\begin{equation}
\epsilon_{oper}=2\xi_L(\underline n_{oper}; \frac{n_{min}}{\underline{n}_{uu}}(1-\frac{n_{min}}{2\underline{n}_{uu}}), \underline{n}_{uu})
\end{equation}
The calculation details of $\underline{n_{01}^\prime}$ and $\underline{n_{10}^\prime}$ are shown in the Appendix~\ref{noper}.

We can then calculate the key rate by:
\begin{equation}\label{opper}
\begin{split}
R=&\frac{1}{N_{tol}}\{{\underline n}_{oper}[1-H({\overline e}_{ph}^\prime)]-fn_{ot}H(E_{OZ})\\
&-(\log_2 \frac{2}{\varepsilon_{sec}}-2\log_2\frac{1}{\sqrt{2}\varepsilon_{PA}\hat{\varepsilon}})\}.
\end{split}
\end{equation}
Here $\overline e_{ph}^\prime=\bar{M}_{oper}/\underline{n}_{oper}$ is the upper bound of the phase-flip error rate after OPER, $n_{ot}$ is the number of survived bits after OPER, and $E_{OZ}$ is the bit-flip error rate after OPER. Moreover, the result can be even better if we take active odd parity pairing which will produce more odd-parity pairs.

\subsection{AOPP-SNS}
We name the SNS protocol~\cite{wang2018twin} with AOPP \cite{xu2020sending,jiang2020zigzag} in the post data processing as AOPP-SNS. In the AOPP, we shall take odd-parity grouping actively so that we can obtain more odd-parity pairs than the OPER does. We divide the odd-parity pairs in AOPP into $g$ subsets so that the number of pairs in each subsets is smaller than the number of odd-parity pairs in OPER. The final key distillation taken on each subset has no difference from that taken in an OPER using partial of its odd-parity pairs. For simplicity, we shall only use two equal subsets here. Say, if we can obtain $2\tilde n_g$ odd-parity pairs by AOPP, we divide these into two subsets, each containing $\tilde n_g$ pairs. We consider the following steps:

1) Take random grouping to bits in set $W$ one by one, we stop grouping at the time $\tilde n_g$ pairs {of} odd-parity pairs are obtained. Suppose $\tilde n_t$ bits are used in the random grouping when $\tilde n_g$ odd-parity pairs are created. This $\tilde n_t$ is an observed number and therefore we don’t have to consider the statistical fluctuation in our calculation. We shall simply use $\tilde n_t = \frac {\tilde n_g n_t^2}{2N_0N_1}$ in our numerical simulation. 

2) Take AOPP to the $n_t$ bits in set $W$. We obtain $2\tilde n_g$ pairs and divide them into two equal subsets. Each subset contains $\tilde n_g$ odd parity pairs {which could have come from OPER.} We can calculate the key rate of exch subset by the formula of Eq.~\eqref{opper} of OPER, in the case that we only use tilde $\tilde n_t$ bits there.

\section{Numerical simulation}
We use the linear model to simulate the observed values with certain experiment devices and certain source parameters~\cite{jiang2019unconditional}. We assume symmetric channel and source parameters between Alice and Bob. The decoy state analysis can be used to calculate the lower bound of the number of untagged bits and the upper bound of their corresponding phase-flip error rate before TWCC. The details of decoy state analysis are shown in the Appendix~\ref{decoy}. The details of how to use data before OPER to estimate the phase errors after OPER are shown in the Appendix~\ref{oper}. By setting the failure probability while calculating the effect of statistical fluctuation as $10^{-20}$, and other failure probabilities as $10^{-20}$, too, we achieve a security level of $1.39\times 10^{-9}$, $2.33\times 10^{-9}$ and $4.66\times 10^{-9}$ in the standard TWCC, OPER and AOPP, respectively. 

\begin{table}[htbp]
\begin{ruledtabular}
\begin{tabular}{cccccc}
$p_d$& $e_d$ &$\eta_d$ & $f$ & $\alpha_f$& $\xi_c$ \\
\hline
$1.0\times10^{-8}$& $3\%$  & $30.0\%$ & $1.1$ & $0.2$ &$10^{-20}$ \\ 
\end{tabular}
\end{ruledtabular}
\caption{List of experimental parameters used in numerical simulations. Here $p_d$ is the dark count rate of Charlie's detectors; $e_d$ is the misalignment-error probability; $\eta_d$ is the detection efficiency of Charlie's detectors; $f$ is the error correction inefficiency; $\alpha_f$ is the fiber loss coefficient ($dB/km$); $\xi_c$ is the failure probability while calculating the effect of statistical fluctuation.}\label{exproperty}
\end{table}

\begin{figure}
\centering
\includegraphics[width=8cm]{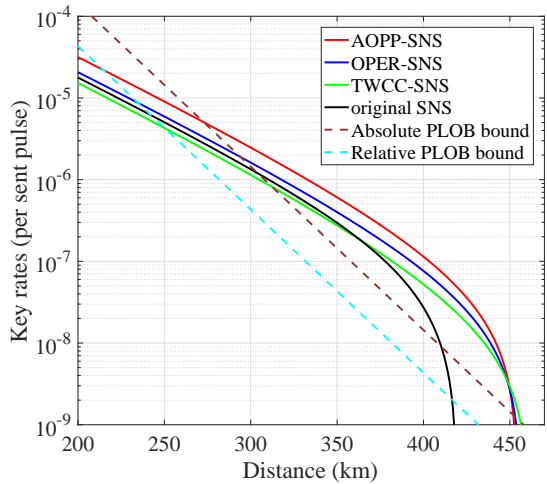}
\caption{The key rates of different protocols. Here we set $N_{tol}=10^{12}$. The other experiment parameters used in the numerical simulation are shown in Table.~\ref{exproperty}. The absolute PLOB bound and the relative PLOB bound are the bound with whatever devices and the practical bound assuming the limited detection efficiency, respectively.}\label{fig20}
\end{figure}

Figure \ref{fig20} are the comparison of the key rates of different protocols. We set $N_{tol}=10^{12}$ in Figure \ref{fig20}. The other experiment parameters used in the numerical simulation are shown in Table.~\ref{exproperty}. We find that with TWCC, the key rate of the SNS protocol of TF-QKD in a large distance range can by far exceed the PLOB bound~\cite{pirandola2017fundamental} as a benchmark of key rate of QKD established by Pirandola, Laurenza, Ottaviani, and Banchi~\cite{pirandola2017fundamental}. The absolute PLOB bound and the relative PLOB bound are the bound with whatever devices and the practical bound assuming the limited detection efficiency, respectively~\cite{pirandola2017fundamental}. Figure \ref{fig20} shows that in the case of finite-key size, the TWCC method can improve the maximum distance of the SNS protocol of TF-QKD by 50 km, and greatly improve the key rate at long distances. The furthest distance of those three improved method:  the standard TWCC, OPER and AOPP, are the same, but in almost all distances, the key rates of AOPP method are the highest.

\begin{figure}
\centering
\includegraphics[width=8cm]{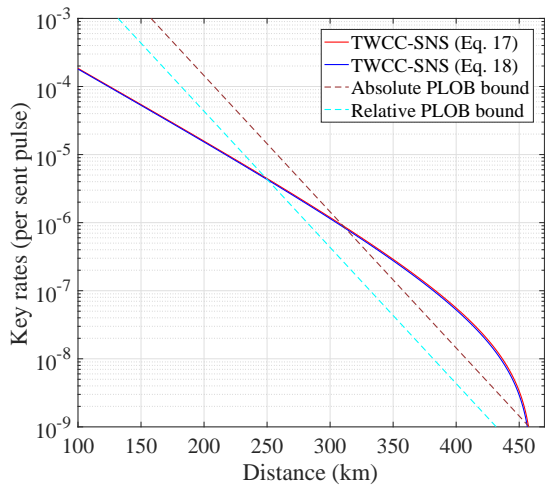}
\caption{The key rates standard TWCC with Eq.~\eqref{mie} and Eq.~{\eqref{strict3}}}\label{twcc2e}
\end{figure}

In the calculation of standard TWCC of Figure \ref{fig20}, we use the Eq.~\eqref{strict3} to estimate the phase errors after error rejection. In Figure \ref{twcc2e}, we compare the key rates of standard TWCC with Eq.~\eqref{mie} and {Eq.~\eqref{strict3}}. The simulation results show that the key rates of those two method are almost the same in all distances.

\section{Conclusion}
In TWCC, the probability of tagging or untagging for each two-bit random groups are not independent. We rigorously solve this problem by imagining a virtual set of bits where every bit is independent and identical. We show that we can naively regard the bits in the real set to be independent and identical and get the bound values by applying Chernoff bound, with the failure probability multiplied $2$. We also show how to apply our mathematical results to the SNS protocol with several TWCC methods. Numerical results show that the TWCC method can improve the maximum distance of the SNS protocol of TF-QKD by 50 km, and greatly improve the key rate at long distances.

\appendix
\section{The proofs}\label{lemma11}
\subsection{The proof of Lemma~\ref{ll1}}
The proof of Lemma \ref{ll1} is very simple. Since we take the grouping {\em randomly}, the outcome probability distribution over numbers of white-white pairs or other kind of pairs is independent of the initial positions of white balls or black balls before grouping, it only depends on the initial number of white balls. Suppose $k_1-k_2=\Delta \ge 0$. If we randomly label any $\Delta$ white balls in set $W_1$ and change them into $\Delta$ black balls, we shall obtain a set equivalent to set $W_2$ for the random grouping process. Also, we may choose to obtain the random grouping result of set $W_2$ by this: we start with set $W_1$, randomly label $\Delta$ white balls before grouping. After grouping, we change those $\Delta$ white balls initially labelled into the black. This shows that by whatever grouping method, a $ww$ pair corresponding to the initial set $W_2$ is always a $ww$ pair corresponding to the initial set $W_1$, but the reverse is not necessarily. This means by whatever grouping method, if the outcome corresponding to initial set $W_2$ satisfies the condition $n_{ww}\ge \underline n_{ww}$, the outcome corresponding to initial set $W_1$ must also satisfy the same condition. This completes Lemma~\ref{ll1} from the ergodic viewpoint of probability.

According to {\bf Notation 2.2}, set $[p_u,N]_{iid}$ can be regarded as the probability distribution over set $[k,N]$. Explicitly, the probability on $[k,N]$ is:
\begin{equation}
\tilde p(k) =C_N^k p_u^k (1-p_u)^{N-k}.
\end{equation}
We shall use this in our proofs of Theorems.
\subsection{The proof of Theorem \ref{th11}}
Starting from the failure probability for $n_{ww} \ge \underline{n}_{ww}$ with the virtual set $[p_u,N]_{iid}$, we have
	\begin{equation*}
	\begin{split}
	&\varepsilon(\underline{n}_{ww}|[p_u,N]_{iid})\\
	=&\sum_{k}\tilde p(k)\varepsilon(\underline{n}_{ww}|[k,N])\\
	=& \sum_{k\le k_{u}} \tilde p(k)\varepsilon(\underline{n}_{ww}|[k,N])+ \sum_{k>k_{u}}\tilde p(k) \varepsilon(\underline{n}_{ww}|[k,N])\\
	\ge& \sum_{k \le k_{u}} \tilde p(k) \varepsilon (\underline{n}_{ww}|[k,N]) \\
	\ge& \sum_{k \le k_{u}} \tilde p(k) \varepsilon(\underline{n}_{ww}|[k_{u},N])\\
	\ge& \gamma_{iid}\varepsilon(\underline{n}_{ww}|[k_{u},N])\\
	\end{split}
	\end{equation*}
We have used Lemma \ref{ll1} in the second inequality above. This ends the proof of Theorem~\ref{th11}.

For the binomial distribution $B(M,p)$, $np$ is its median if $Mp$ is a integer. Thus by setting $p_u=k_u/N$ and combining with Lemma~\ref{ll1}, we immediately transform Theorem~\ref{th11} to Result 2.1.

\subsection{The proof of Result {4.1}}
To proof Result 4.1, we first introduce the following lemma and theorem.

\begin{lemma}\label{ll2}
		For sets  $W_1=[k_1,N]$, $W_2=[k_2, N]$, the inequality
	\begin{equation}
	\label{lemin3}
	\varepsilon(\underline {n}_{wb}+1|W_2)\ge \varepsilon(\underline {n}_{wb}|W_1)
	\end{equation}
	always hold for whatever non-negative integer $\underline n_{wb}$ provided that $ 0\le k_2 \le k_1\le N/2$;
\end{lemma}

We shall prove Lemma~\ref{ll2} in two cases. 

For the case $ 0\le k_2 <\underline {n}_{wb}\le k_1\le N/2$, it is easy to check 
\begin{equation}
\varepsilon(\underline {n}_{wb}+1|W_2)=1\ge \varepsilon(\underline {n}_{wb}|W_1)
\end{equation}. 

For the case $ 0\le \underline {n}_{wb}\le k_2 \le k_1\le N/2$, if 
\begin{equation}\label{pr1}
\varepsilon(2\underline{n}_e+1|[2k,2n])\ge\varepsilon(2\underline{n}_e+1|[2k+1,2n]),
\end{equation}
and
\begin{equation}\label{pr2}
\varepsilon(2\underline{n}_e+1|[2k,2n])\ge\varepsilon(2\underline{n}_e+1|[2k+2,2n]),
\end{equation}
hold for any $\underline{n}_e$ and $k$ that satisfied $2\underline{n}_e+1 < 2k \le n-1$, then we obviously have $\varepsilon(\underline {n}_{wb}+1|W_2)\ge \varepsilon(\underline {n}_{wb}|W_1)$. Our task now is reduced to prove Eqs.~\eqref{pr1} and \eqref{pr2}.

Denote $P(2l)_{2k}$ as the probability that there are $2l$ $wb$-pairs after performing random grouping to set $[2k,2n]$, and we have
\begin{equation}
P(2l)_{2k}=\frac{C_n^{k-l}C_{n-(k-l)}^{2l}A_{2k}^{2k}A_{2n-2k}^{2n-2k}2^{2l}}{A_{2n}^{2n}},
\end{equation}  
where $C_a^b$ is the number of combinations and $A_a^b$ is the number of arrangements. It is easy to check that
\begin{align}\label{eq15}
P(2l+1)_{2k+1}=&\frac{2n-2k-2l}{2n-2k}P(2l)_{2k}\nonumber\\
&+\frac{2l+2}{2n-2k}P(2l+2)_{2k},\\
P(2l)_{2k+2}=&\frac{2n-(2k+1)-(2l-1)}{2n-(2k+1)}P(2l-1)_{2k+1}\nonumber \\
&+\frac{2l+1}{2n-(2k+1)}P(2l+1)_{2k+1}.
\end{align}

According to \textbf{Notation 2.4}, we have
\begin{equation}\label{eq16}
\begin{split}
&\varepsilon(2\underline{n}_e+1|[2k,2n])=\sum_{l=0}^{\underline{n}_e}P(2l)_{2k},\\
&\varepsilon(2\underline{n}_e+1|[2k+1,2n])=\sum_{l=1}^{\underline{n}_e}P(2l-1)_{2k+1},\\
&\varepsilon(2\underline{n}_e+1|[2k+2,2n])=\sum_{l=0}^{\underline{n}_e}P(2l)_{2k+2}.
\end{split}
\end{equation}
Combining Eq.~\eqref{eq15} and \eqref{eq16}, we have
\begin{equation}
\begin{split}
&\varepsilon(2\underline{n}_e+1|[2k+1,2n])=\varepsilon(2\underline{n}_e+1|[2k,2n])-x_1,\\
&\varepsilon(2\underline{n}_e+1|[2k+2,2n])=\varepsilon(2\underline{n}_e+1|[2k,2n])-x_1+x_2,
\end{split}
\end{equation}
where
\begin{equation}
\begin{split}
&x_1=\frac{2n-2k-2\underline{n}_e}{2n-2k}P(2\underline{n}_e)_{2k},\\
&x_2=\frac{2\underline{n}_e+1}{2n-(2k+1)}P(2\underline{n}_e+1)_{2k+1}.
\end{split}
\end{equation}

As $x_1$ is a positive number, thus Eq.~\eqref{pr1} holds. As $2k\le n-1$, we have
\begin{equation}
\frac{x_1}{x_2}=\frac{2n}{2k+1}-1\ge 1,
\end{equation}
thus Eq.~\eqref{pr2} holds.

This ends the proof of Lemma~\ref{ll2}.

With Lemma~\ref{ll2},we have
\begin{theorem}\label{th22}
	The inequality
	\begin{equation}\label{nne}
	 \varepsilon(\underline{n}_{wb}|[k_u,N]) \le  \frac {\varepsilon(\underline{n}_{wb}+1|[p_u,N]_{iid})}{\gamma_{iid}^{\prime}}
	\end{equation}
	always holds with whatever nature number $\underline{n}_{wb}$, $k_{u}\le\frac{N}{2}$ and whatever probability value $p_u$.
	Here
	\begin{equation}
	\begin{split}
	\varepsilon(\underline{n}_{wb}+1|[p_u,N]_{iid})=&\sum_{k=0}^{\underline{n}_{wb}}C_{\frac{N}{2}}^k\mathcal P^k(1-\mathcal P)^{\frac{N}{2}-k}
	\end{split}
	\end{equation}
	where $\mathcal P = 2p_u(1-p_u)$ and
	\begin{equation}
	\gamma_{iid}^{\prime}=\sum_{k=0}^{k_u}\tilde{p}(k).
	\end{equation}
	
\end{theorem}

According to the definition of $\varepsilon(\underline{n}_{wb}+1|[p_u,N]_{iid})$, we have
\begin{align*}
&\varepsilon(\underline{n}_{wb}+1|[p_u,N]_{iid})\\
&=\sum_{k}\tilde{p}(k)\varepsilon(\underline{n}_{wb}+1|[k,N])\\
&\ge\sum_{k\le k_u}\tilde{p}(k)\varepsilon(\underline{n}_{wb}+1|[k,N])\\
&\ge\sum_{k\le k_u}\tilde{p}(k)\varepsilon(\underline{n}_{wb}|[k_u,N])\\
&=\gamma_{iid}^{\prime} \varepsilon(\underline{n}_{wb}|[k_u,N]).
\end{align*}
Here we use Lemma~\ref{ll2} for the second inequality. This ends the proof of Theorem \ref{th22}.

For the binomial distribution $B(M,p)$, $Mp$ is its median if $Mp$ is a integer. Thus by setting $p_u=k_u/N$ and combining with Lemma~\ref{ll2}, we immediately transform Theorem~\ref{th22} to Result 4.1.

\section{The decoy state analysis}\label{decoy}
Since the original SNS protocol was proposed \cite{wang2018twin}, it has been further studied extensively~\cite{yu2019sending,hu2019general,xu2020sending,jiang2020zigzag}. The 4-intensity and 3-intensity SNS protocols with weak coherent state (WCS) sources are usually applied in the experiment. In the 4-intensity SNS protocol~\cite{yu2019sending,jiang2019unconditional}, there are four sources with intensities $0,\mu_{a1},\mu_{a2}$ and $\mu_{az}$ at Alice's side and intensities $0,\mu_{b1},\mu_{b2}$ and $\mu_{bz}$ at Bob's side. If we set $\mu_{a2}=\mu_{az}$ and $\mu_{b2}=\mu_{bz}$, the 4-intensity SNS protocol becomes the 3-intensity protocol. In this paper, we take the 4-intensity SNS protocol as an example to show our calculation method.

In the whole protocol, Alice and Bob (\emph{they}) send $N$ pulse pairs to Charlie, who is assumed to perform interferometric measurements on the received pulses and announces the measurement results to \emph{them}. If only one detector clicks, \emph{they} would take it as an one-detector heralded event. At each time window, Alice (Bob) randomly decides whether it is a decoy window with probability $1-p_{z}$, or a signal window with probability $p_{z}$. If it is a signal window, with probability $\epsilon_A$ ($\epsilon_B$), Alice (Bob) prepares a  pulse with intensity $\mu_{az}$ ($\mu_{bz}$), and denote it as bit $1$ ($0$); with probability $1-\epsilon_a$ ($1-\epsilon_b$), Alice (Bob) prepares a vacuum pulse, and denote it as bit $0$ ($1$). If it is a decoy window, Alice (Bob) randomly prepares a vacuum pulse or a pulse with state $\ket{e^{i\theta_{a1}}\sqrt{\mu_{a1}}}$ or $\ket{e^{i\theta_{a2}}\sqrt{\mu_{a2}}}$ ($\ket{e^{i\theta_{b1}}\sqrt{\mu_{b1}}}$ or $\ket{e^{i\theta_{b2}}\sqrt{\mu_{b2}}}$) with probabilities $p_{a0}=1-p_{a1}-p_{a2}$, $p_{a1}$ and $p_{a2}$, ($p_{b0}=1-p_{b1}-p_{b2}$, $p_{b1}$ and $p_{b2}$) respectively, where $\theta_{a1},\theta_{a2},\theta_{b1}$ and $\theta_{b2}$ are different in different windows, and are random in $[0,2\pi)$. We set the following constraint for the security of SNS protocol~\cite{hu2019general}
\begin{equation}\label{securec}
\frac{\mu_{a1}}{\mu_{b1}}=\frac{\epsilon_{a}(1-\epsilon_b)\mu_{az}e^{-\mu_{az}}}{\epsilon_b(1-\epsilon_a)\mu_{bz}e^{-\mu_{bz}}}.
\end{equation}    
For the symmetric SNS protocol, saying $p_{az}=p_{bz},p_{a0}=p_{b0},\mu_{az}=\mu_{bz}$ and so on, the constraint \eqref{securec} is automatically satisfied.

After \emph{they} repeat the above process for $N$ times, they acquire a series of data. For a time window that both \emph{them} decide a signal window, it is a $Z$ window. The one-detector heralded events in $Z$ windows are effective events, and the corresponding bits of those effective events formed the $n_t$-bit raw key strings, which are used to extract the final keys. For a time window that both \emph{them} decide send out a pulse with intensities $\mu_{a1}$ and $\mu_{b1}$ respectively, and their phases satisfy
\begin{equation}
1-|\cos{(\theta_{a1}-\theta_{b1})}|\le \lambda,
\end{equation}
where $\lambda$ is a small positive number, it is an $X$ window. The one-detector heralded events in $X$ windows are effective events. And for an effective event in the $X$ window, if $\cos{(\theta_{a1}-\theta_{b1})}>0$ and Charlie announces a click of right or $\cos{(\theta_{a1}-\theta_{b1})}<0$ and Charlie announces a click of left, it is defined as an error effective event. The effective events in $X$ windows are used to estimate the phase-flip error rate. And $\lambda$ would be taken as an optimized parameter to get the best estimation of phase-flip error rate.

We denote the vacuum source, the WCS source with intensity $\mu_{a1},\mu_{a2}$, and $\mu_{az}$ ($\mu_{b1},\mu_{b2}$, and $\mu_{bz}$) of Alice (Bob) by $ao,ax,ay$ and $az$ ($bo,bx,by$, and $bz$). We simplify the symbols of  two pulse sources $a\kappa,b\zeta(\kappa,\zeta=o,x,y$ as $\kappa\zeta$. We denote the number of pulse pairs of source $\kappa\zeta$ sent out in the whole protocol by $N_{\kappa\zeta}$, and the total number of one-detector heralded events of source $\kappa\zeta$ by $n_{\kappa\zeta}$. We define the counting rate of source $\kappa\zeta$ by $S_{\kappa\zeta}=n_{\kappa\zeta}/N_{\kappa\zeta}$, and the corresponding expected value by $\mean{S_{\kappa\zeta}}$. The Chernoff bound can be used to estimate the lower and upper bound of the expected values according to their observed values.

Then we can use the decoy-state method to calculate the lower bounds of the expected values of the counting rate of single-photon states $\oprod{01}{01}$ and $\oprod{10}{10}$, which are~\cite{hu2019general}
\begin{align}
\label{s01mean}\mean{\underline{s_{01}}}&= \frac{\mu_{b2}^{2}e^{\mu_{b1}}\mean{S_{ox}}-\mu_{b1}^{2}e^{\mu_{b2}}\mean{S_{oy}}-(\mu_{b2}^{2}-\mu_{b1}^{2})\mean{S_{oo}}}{\mu_{b2}\mu_{b1}(\mu_{b2}-\mu_{b1})},\\
\mean{\underline{s_{10}}}&= \frac{\mu_{a2}^{2}e^{\mu_{a1}}\mean{S_{xo}}-\mu_{a1}^{2}e^{\mu_{a2}}\mean{S_{yo}}-(\mu_{a2}^{2}-\mu_{a1}^{2})\mean{S_{oo}}}{\mu_{a2}\mu_{a1}(\mu_{a2}-\mu_{a1})}.
\end{align}
Then we can get the lower bound of the expected value of the counting rate of untagged photons
\begin{equation}
\mean{\underline{s_1}}=\frac{\mu_{a1}}{\mu_{a1}+\mu_{b1}}\mean{\underline{s_{10}}}+\frac{\mu_{b1}}{\mu_{a1}+\mu_{b1}}\mean{\underline{s_{01}}},
\end{equation}
and the lower bounds of the expected values of the the untagged bits $\mean{\underline{n_1}}$, untagged bits $1$, $\mean{\underline{n_{10}}}$, and untagged bits $0$, $\mean{\underline{n_{01}}}$
\begin{align}
\mean{\underline{n_{1}}}=&Np_{az}p_{bz}[\epsilon_a(1-\epsilon_b)\mu_{az}e^{-\mu_{az}}\nonumber \\
&+\epsilon_b(1-\epsilon_a)\mu_{bz}e^{-\mu_{bz}}]\mean{\underline{s_{1}}},\\
\mean{\underline{n_{10}}}=&Np_{az}p_{bz}\epsilon_a(1-\epsilon_b)\mu_{az}e^{-\mu_{az}}\mean{\underline{s_{10}}},\\
\mean{\underline{n_{01}}}=&Np_{az}p_{bz}\epsilon_b(1-\epsilon_a)\mu_{bz}e^{-\mu_{bz}}\mean{\underline{s_{10}}}.
\end{align}
With Chernoff bound, we can estimate the lower bounds of the number of untagged bits $1$, ${\underline{n_{10}}}$, and untagged bits $0$, ${\underline{n_{01}}}$ 
\begin{equation}\label{n11111}
\underline{n_{10}}=\varphi^L(\mean{\underline{n_{10}}}),\underline{n_{01}}=\varphi^L(\mean{\underline{n_{01}}}),\underline{n_{1}}=\underline{n_{01}}+\underline{n_{10}},
\end{equation}
where $\varphi^L(x)$ are the lower bound while using Chernoff bound to estimate the real value according to the expected value.

We denote the number of total pulses sent out in the $X$ windows by $N_{X}$, and the number of error effective events by $m_{X}$, then we have the error counting rate of $X$ windows
\begin{equation}
T_{X}=\frac{m_{X}}{N_{X}}.
\end{equation}
Then we have
\begin{equation}\label{e1}
\mean{\overline{e_1^{ph}}}=\frac{\mean{T_{X}}-e^{-\mu_{a1}-\mu_{b1}}\mean{S_{oo}}/2}{e^{-\mu_{a1}-\mu_{b1}}(\mu_{a1}+\mu_{b1})\mean{\underline{s_1}}},
\end{equation}
where $\mean{T_{X}}$ is the expected value of $T_{X}$. Here we have used the fact that the expected value of the error rate of vacuum pulses are always $\frac{1}{2}$.  

Finally, by using Chernoff bound~\cite{chernoff1952measure}, we can get the upper bound of the number of phase-flip errors before TWCC.

\section{Zigzag approach to phase error after OPER}\label{oper}
Here we review the main idea of the zigzag approach~\cite{jiang2020zigzag} on how to calculate the phase-error rate after OPER, with finite data size.

Suppose they have $n_t$ effective bits in $Z$ basis, where $N_u$ of them are untagged bits before OPER. After random pairing, there are $n_{uu}$ untagged pairs, formed by $M=2 n_{uu}$ untagged bits. Given the lower bound number of untagged bits in $Z$ basis, $n_{uu}$ can be lower bounded by our Result 2.1.

For clarity, we image to replace those $N_u$ untagged bits by $N_u$ virtual bipartite entangled single-photons shared by Alice and Bob. However, since there is no bit flip error, each photon lives in a two-dimensional space only. We shall simple call this bipartite entangled single photons by qubits.

Consider those $M=2n_{uu}$ qubits that form the $n_{uu}$ untagged pairs. Before random pairing, they are a random subset of set from all those $N_u$ untagged qubits (recall our Remark 3.1). Therefore we can apply the quantum de Finetti theorem~\cite{renner2005security,renner2007symmetry}. 

{\bf Main idea}: We shall consider the mathematical properties of density operator of those $M$ qubits, $\rho$. According to the quantum de Finetti theorem, there exists another density operator $\tilde \rho$ which has a very small trace distance with $\rho$. We name this $\tilde \rho$ as the associate state of $\rho$. {Among the $M$ qubits for state  $\tilde \rho$, there are $M-r$ qubits in classical mixture of states where every qubit is identical}. We denote set $I_D$ for these $M-r$ qubits.
 Without any loss of generallity, the density operator of qubits in set $I_D$ can be weitten in the following form:
\begin{equation}
	\label{qdef}
	\tilde \rho' = \int_0^{1} f(p)\sigma_p^{\otimes (M-r)} dp
\end{equation}
where $\sigma_p$ is a qubit density operator that has robability $p$ taking a phase error and
$f(p)$ is the probability distribution on phase-error probability $p$. {The above form of state $\tilde \rho'$ means that every qubit is in a certain idependent and identical state $\sigma_p$, and   there is a probability distribution for all possible $\sigma_p$.}
 Although we are not able to calculate the upper bound of  phase error after OPER with the input state $\rho$, we can upper bound the number of phase errors after OPER with the associate state $\tilde \rho$. This also upper bounds the number of phase errors with input $\rho$, with a small failure probability since the trace distance of state $\rho$ and $\tilde \rho$ is very small.

1) For state $\rho$, the number of phase errors $m_e$ is upper bounded by $\bar m_e$, i.e.
\begin{equation}\label{stph}
m_e\le \bar m_e
\end{equation}
with a failure probability at most $\epsilon_1$. This fact is verified by the error test in $X$ basis and the decoy-state analysis.

2) Applying the quantum de Finetti theorem, there exists another $M-$qubit density operator $\tilde \rho$ with the following two mathematical properties: 
i) The trace distance between $\rho$ and $\tilde \rho$ is at most $\epsilon_2$, i.e.
\begin{equation}\label{traceD}
D(\rho,\tilde\rho) \le \epsilon_2
\end{equation}
and ii) In state $\tilde \rho$, $M-r$ qubits are in the classical mixture of independently identically distribution (iid) states, as shown in Eq.~\eqref{qdef}. We denote set $I_D$ for these $M-r$ qubits. Also, Eqs.(\ref{stph}) and (\ref{traceD}) mathematically constraint the number of phase errors $\tilde m_e$ for state $\tilde \rho$ by
\begin{equation}\label{tnphe}
\tilde m_e\le \bar m_e
\end{equation}
with a failure probability $\epsilon_3$.

3) Given properties above for density operator $\tilde \rho$, the number of phase errors $\tilde m_e'$ of set $I_D$ are at most $\bar m_e$, i.e.,
\begin{equation}\label{3e}
\tilde m_e' \le \bar m_e
\end{equation} with a failure probability upper bounded by $\epsilon_3$ . 

4)  
{We can upper bound the value of $p$ in state of Eq.(\ref{qdef}) with failure probability  $\epsilon_4$}. The qubits in set $I_D$ are the classical mixture  states where every qubit has the same probability $p$ to carry a phase-flip error as shown in Eq.(\ref{qdef}). Naviely speaking, this value $p\le 1$. However, we can have a nontrivial upper bound for the value $p$ by applying the constraint of Eq.(\ref{3e}).  Say, if we choose the upper bound value to be $p_e$, we can compute the failure probability 
	$\epsilon_4$ for the inequality
\begin{equation}\label{3pe3}
	p \le  p_e.
\end{equation}
Here in our numerical calculation simulation, we have taken $p_e=\frac{\bar m_e}{M-r}$ as the upper bound of phase-flip error rate of set $I_D$. As shown in the end of this subsection, the failure probability for inequality  (\ref{3pe3}) $\epsilon_4$ is bounded by
\begin{equation}\label{eps4}
	\epsilon_4 \le 2 \epsilon_3
\end{equation}

5) {Number of phase errors after OPER.} Define a pair containing two qubits from set $I_D$ as an $DD$-pair. With step 4), we can regard every qubit in set $I_D$ has independent and  identical probability $p$ to be a phase error with constraint $p\le p_e$. Taking the worst case $p=p_e$ we can upper bound $m_s$, the number of phase error odd-parity $DD$-pairs by
\begin{equation}\label{5ph}
m_s \le  \bar{m}_s
\end{equation}
with a failure probability at most $\epsilon_5$, {where $m_s$ is the number of phase errors after OPER and upper bounded by $\bar{m}_s$.} To explicitly calculate $\bar{m}_s$, we need use the parity check operator given in Ref.~\cite{jiang2020zigzag}.
{In the $M$-qubit associate state $\tilde\rho$, there are $r$ qubits not beloning set $I_D$. Consider the worst case for those $r$ bits not in set $I_D$ in random pairing in OPER : they paticipate in $r$ odd-parity pairs and each pair produces a phase error in its survived bit.} We conclude the final equation for the number of phase-error untagged odd-parity pairs after OPER:
\begin{equation}\label{eodd}
m_{s}^\prime \le  \bar{m}_s+r.
\end{equation} 

6) Eq.(\ref{eodd}) also makes the upper bound of phase errors $m_{odd}$ of survived bits from odd-parity pairs with input of state $\rho$,  
\begin{equation}
m_{odd} \le \bar{m}_s+r.
\end{equation}
with failure probability $\epsilon_6$.

Since all operations are done in $Z$ basis, it makes no difference if each side does the local measurement in the beginning. In this case, it is just a protocol taking random pairing on classical bits. In a protocol with pre-shared single-photon entangled states, the number of odd parity pairs are directly observed. In a real protocol with coherent states from each sides, the the lower bound of number of odd parity untagged pairs can be verified by the decoy-state method and the results in this paper. 
 
All those $\epsilon_i$ are computable. $\epsilon_1$ is done by phase error estimation in the decoy-state method. $\epsilon_2$ is determined by the size of the whole set of untagged bits in $Z$ basis and the value $M$, number of untagged bits for those $n_{uu}$ untagged pairs after random pairing. Lower bound $M$ can be verified by Result 2.1. $\epsilon_3$ is determined by $\epsilon_1$ and $\epsilon_2$ { while $\epsilon_4$ is upper bounded by $2\epsilon_3$ in Eq.(\ref{eps4}).} $\epsilon_5$ is the failure probability of a binomial distribution as shown in Ref.~\cite{jiang2020zigzag}. $\epsilon_6$ is determined by { $\epsilon_5$ and the trace distance between $\rho$ and $\tilde \rho$.}

{Proof of Eq.(\ref{eps4}): Consider Eq.(\ref{qdef}), the failure probability $\epsilon_4$ for inequality (\ref{3pe3}) is 
\begin{equation}\label{fpt1}
	\epsilon_4 = \int_{p_e}^1 f(p) dp.
\end{equation}
To upper bound this $\epsilon_4$, we introduce a notation first first:
\\{\bf Notation C.1} We denote
$\mathcal P (\bar m_e|\Omega)$ for the probability that the $(M-r)$-qubit state $\Omega$ produce more than $\bar m_e$ {phase-flip} errors.
\\ Given this, for the state of qubits in set $I_D$, we can  formulate
\begin{equation}\label{cons3e}
	\epsilon_3 \ge \mathcal P (\bar m_e|\tilde\rho')
\end{equation}
by the constraint for failure probability of Eq.(\ref{3e}) and also
\begin{equation}
	\begin{split}
		\mathcal P (\bar m_e|\tilde\rho') &
		= \int_0^{1} f(p)\mathcal  P(\bar m_e|\sigma_p^{\otimes (M-r)}) dp\\
		& =\int_0^{p_e} f(p)\mathcal P(\bar m_e|\sigma_p^{\otimes (M-r)}) dp \\&+ \int_{p_e}^{1} f(p)\mathcal P(\bar m_e|\sigma_p^{\otimes (M-r)}) dp\\
		& \ge \int_{p_e}^{1} f(p)\mathcal P(\bar m_e|\sigma_{p_e}^{\otimes (M-r)}) dp\\ & \ge
		\frac{1}{2}\int_{p_e}^{1} f(p) dp
	\end{split}
\end{equation}
where we have {used} $p_e=\frac{\bar m_e}{M-r}$ {and the fact that $\mathcal P(\bar m_e|\sigma_{p_e}^{\otimes (M-r)})$} is a rising functional of $p$ for the last inequality above. Consider the definition of $\epsilon_4$ in Eq.(\ref{fpt1}) and also the constraint in Eq.(\ref{cons3e}) above, we immediately conclude Eq.(\ref{eps4}).}
\section{The calculation method of  $\underline{n}_{oper}$}\label{noper}
Recall that set $Y$ containing $2\underline{n}_{uu}$ bits that formed $\underline{n}_{uu}$ survived untagged pairs after OPER. Bits in set $Y$ are randomly chosen from the initial $\underline{n_1}$ untagged bits which containing $\underline{n_{01}}$ untagged bits $0$ and $\underline{n_{01}}$ untagged bits $0$ before OPER. Thus $n_{01}^\prime$ and $n_{10}^\prime$, the numbers of untagged bits $0$ and untagged bits $1$ in set $Y$ satisfy the hypergeometric distribution $Hy(\underline{n_{01}},\underline{n_1},2\underline{n}_{uu})$ and $Hy(\underline{n_{10}},\underline{n_1},2\underline{n}_{uu})$ respectively, where $Hy(K,N,n)$ is the hypergeometric distribution that perform $n$ draws, without replacement, from a finite population of size $N$ that contains $K$ target objects. With the tail bounds of hypergeometric distribution~\cite{hoeffding1994probability}, we have
\begin{equation}
\underline{n_{01}^\prime}=2\underline{n}_{uu}(\frac{\underline{n_{01}}}{\underline{n_{1}}}-\sqrt{\frac{-\ln{\xi_h}}{2\underline{n}_{uu}}}),
\end{equation} 
and
\begin{equation}
\underline{n_{10}^\prime}=2\underline{n}_{uu}(\frac{\underline{n_{10}}}{\underline{n_{1}}}-\sqrt{\frac{-\ln{\xi_h}}{2\underline{n}_{uu}}}),
\end{equation} 
where $\xi_h$ is the failure probability.

It is easy to check the worst case of $\underline{n}_{oper}$ is achieved in the lower bounds of $n_{01}^\prime$ or $n_{10}^\prime$. Let $n_{min}=\min(\underline{n_{01}^\prime},\underline{n_{10}^\prime})$. If $n_{min}=\underline{n_{01}^\prime}$, relate bits $0$ in set $Y$ to white balls and bits $1$ to black balls, and if $n_{min}=\underline{n_{10}^\prime}$, relate bits $1$ in set $Y$ to white balls and bits $0$ to black balls, we have $n_{oper}\ge\underline{n}_{oper}$ with a failure probability $\epsilon_{oper}$
and   
\begin{equation}
\epsilon_{oper}=2\xi_L(\underline n_{oper}; \frac{n_{min}}{\underline{n}_{uu}}(1-\frac{n_{min}}{2\underline{n}_{uu}}), \underline{n}_{uu})
\end{equation}

\bibliography{refs}

\end{document}